
%
%
%
%
\ifx\epsffile\undefined\message{(FIGURES WILL BE IGNORED)}
\def\insertplot#1#2{}
\def\insertfig#1#2{}
\else\message{(FIGURES WILL BE INCLUDED)}
\def\insertplot#1#2{
\midinsert\centerline{{#1}}\vskip0.2truein\centerline{{\epsfxsize=4.0truein
\epsffile{#2}}}\vskip0.5truecm\endinsert}
\def\insertfig#1#2{
\midinsert\centerline{\epsffile{#2}}\centerline{{#1}}\endinsert}
\fi
\input harvmac
%
%
%
%
\ifx\answ\bigans
\else
\hoffset=-0.25truein
\voffset=-0.50truein
\output={
  \almostshipout{\leftline{\vbox{\pagebody\makefootline}}}\advancepageno
}
\fi
%
%
%

%
%

%
%
\def\UCSD#1#2{\noindent#1\hfill #2%
\bigskip\supereject\global\hsize=\hsbody%
\footline={\hss\tenrm\folio\hss}}
%
%
\def\abstract#1{\centerline{\bf Abstract}\nobreak\medskip\nobreak\par #1}
%
%
%
%
\edef\tfontsize{ scaled\magstep3}
 \tfontsize  \tfontsize
 \tfontsize \font\titlei=cmmi10 \tfontsize
\font\titleis=cmmi7 \tfontsize \font\titleiss=cmmi5 \tfontsize
\font\titlesy=cmsy10 \tfontsize \font\titlesys=cmsy7 \tfontsize
\font\titlesyss=cmsy5 \tfontsize  \tfontsize
\skewchar\titlei='177 \skewchar\titleis='177 \skewchar\titleiss='177
\skewchar\titlesy='60 \skewchar\titlesys='60 \skewchar\titlesyss='60
%
%
%
%
%
\def\inv{^{\raise.15ex\hbox{${\scriptscriptstyle -}$}\kern-.05em 1}}
\def\lbar{{\lower.35ex\hbox{$\mathchar'26$}\mkern-10mu\lambda}} 

%
%
%
%
\def\slash#1{\rlap{$#1$}/} 
\def\dsl{\,\raise.15ex\hbox{/}\mkern-13.5mu D} 
\def\delsl{\raise.15ex\hbox{/}\kern-.57em\partial}
\def\Ksl{\hbox{/\kern-.6000em\rm K}}
\def\Asl{\hbox{/\kern-.6500em \rm A}}
\def\Dsl{\hbox{/\kern-.6000em\rm D}} 
\def\Qsl{\hbox{/\kern-.6000em\rm Q}}
\def\gradsl{\hbox{/\kern-.6500em$\nabla$}}
%
%
\def\lspace{\ifx\answ\bigans{}\else\qquad\fi}
\def\lbspace{\ifx\answ\bigans{}\else\hskip-.2in\fi} 
%
%
\def\boxeqn#1{\vcenter{\vbox{\hrule\hbox{\vrule\kern3pt\vbox{\kern3pt
        \hbox{${\displaystyle #1}$}\kern3pt}\kern3pt\vrule}\hrule}}}
%
%
\def\mbox#1#2{\vcenter{\hrule \hbox{\vrule height#2in
\kern#1in \vrule} \hrule}}
%
%
%
%
\def\CA{{\cal A}}   \def\CD{{\cal D}}
   
   \def\CL{{\cal L}}
  \def\CO{{\cal O}}

%
%
%
%
%

%

\def\bar#1{\overline{#1}}

\def\abs#1{\left| #1\right|}

\def\darr#1{\raise1.5ex\hbox{$\leftrightarrow$}\mkern-16.5mu #1}

%
%
\def\half{{\textstyle{1\over2}}} 
\def\frac#1#2{{\textstyle{#1\over #2}}} 
%
%
%
%
\def\tr{\mathop{\rm tr}}
\def\Tr{\mathop{\rm Tr}}

\def\Re{\mathop{\rm Re}}

\def\MeV{{\rm MeV}}

%
%
%
%

%
%
\def\ltap{\ \raise.3ex\hbox{$<$\kern-.75em\lower1ex\hbox{$\sim$}}\ }
\def\gtap{\ \raise.3ex\hbox{$>$\kern-.75em\lower1ex\hbox{$\sim$}}\ }
\def\gl{\ \raise.5ex\hbox{$>$}\kern-.8em\lower.5ex\hbox{$<$}\ }
\def\roughly#1{\raise.3ex\hbox{$#1$\kern-.75em\lower1ex\hbox{$\sim$}}}
%
%
        \def\etc{\hbox{\it etc.}}
        
\def\etal{\hbox{\it et al.}}

\def\np#1#2#3{Nucl. Phys. B{#1} (#2) #3}
\def\pl#1#2#3{Phys. Lett. {#1}B (#2) #3}
\def\prl#1#2#3{Phys. Rev. Lett. {#1} (#2) #3}
\def\physrev#1#2#3{Phys. Rev. {#1} (#2) #3}

\relax

\def\lcsb{\Lambda_{\chi}}
\def\lng{\Lambda\rightarrow n\gamma}
\def\spg{\Sigma^+\rightarrow p\gamma}
\def\sng{\Sigma^0\rightarrow n\gamma}
\def\xmsg{\Xi^-\rightarrow\Sigma^-\gamma}
\def\xsg{\Xi^0\rightarrow\Sigma^0\gamma}
\def\xlg{\Xi^0\rightarrow\Lambda\gamma}
\def\ds{\Delta S = 1}
\def\dfour#1{{d^4 #1\over (2\pi)^4}}
\def\ai{{\rm Im}\ a}
\vskip 1.in
\centerline{{\titlefont{Weak Radiative Hyperon Decays}}}
\medskip
\centerline{{\titlefont{in Chiral Perturbation Theory}}}
\vskip .5in
\centerline{Elizabeth Jenkins,${}^a$\footnote{${}^*$}{On leave from the
University of California at San Diego} Michael Luke,${}^b$}
\smallskip
\centerline{Aneesh V.~Manohar,${}^{a\,*}$ and
Martin J. Savage${}^b$\footnote{$^{\dagger}$}{SSC Fellow}}
\bigskip
\centerline{\sl a) CERN TH Division, CH-1211 Geneva 23, Switzerland}
\centerline{\sl b) Department of Physics, University of California at San
Diego,}\centerline{\sl 9500 Gilman Drive, La Jolla, CA 92093}
\vfill
\abstract{The parity-conserving $a$ and parity-violating $b$
amplitudes for weak radiative hyperon decay are studied using chiral
perturbation theory. The imaginary parts of $a$ and $b$ are computed
using unitarity.  The real part of $b$ is dominated by a one-loop
infrared divergent graph which is computed. The
real part of $a$ has a large theoretical uncertainty and
cannot be calculated reliably.  Counterterms for
the $a$ and $b$ amplitudes are classified using $CPS$ symmetry.
The experimental values for decay widths and asymmetries
are consistent with theory, with the exception of the asymmetry
parameter for the $\spg$ decay.
}
\vfill
\UCSD{\vbox{
\hbox{CERN-TH.6690/92}
\hbox{UCSD/PTH 92-33}
\hbox{hep-ph/9210265}}}{October 1992}
\eject

\newsec{Introduction}

Weak radiative hyperon decays (WRHD), such as $\lng$, $\spg$, \etc, have
been of theoretical and experimental interest for some time.
Early theoretical work on these decays focused on pole
models, in which the initial baryon turns into an intermediate state
by a $\ds$ weak transition, followed by the decay of the intermediate
state by the radiation of a photon \ref\poleref{R.H. Graham and S. Pakvasa,
\physrev{140}{1965}{B1144}\semi M.B. Gavela \etal, \pl{101}{1981}{417}\semi
F. Close and H.R. Rubinstein, \np{173}{1980}{477}}\ref\farrar{G. Farrar,
\physrev{D4}{1971}{212}}.  The limited success of pole models for
WRHD led to a variety of other approaches.  Gilman and Wise
\ref\gw{F.J. Gilman and M.B. Wise, \physrev{D19}{1979}{976}}
investigated the possibility that WRHD proceeds via a direct
$s\rightarrow d\gamma$ transition.  The authors concluded that
this assumption was not correct, however, because the theoretical
predictions of the model were incompatible with experiment.  Further
attempts include the study of WRHD in the Skyrme model \ref\kao{W.F. Kao
and H.J. Schnitzer, \physrev{D37}{1988}{1912}}\ and using an effective
Lagrangian \ref\cohen{A. Cohen, \pl{160}{1985}{177}}.

Although no
theoretical analysis has resulted in a complete description of WRHD thus
far,
certain long distance contributions to WRHD have been determined using very
general arguments.
Since the decay $\lng$ can proceed via the physically allowed weak decay
$\Lambda\rightarrow p\pi^-$, followed by $p \pi^-\rightarrow n\gamma$, the
imaginary part of the $\lng$ amplitude is determined in terms of the known
amplitudes for the hyperon nonleptonic decay $\Lambda\rightarrow p\pi^-$
and for pion photoproduction $p\pi \rightarrow \gamma n$.
This method was used by Farrar \farrar\ to place a unitarity
lower bound on WRHD rates.  Kogan and Shifman \ref\kogan{Ya. I. Kogan and
M.A. Shifman, Sov. J. Nucl. Phys. 38 (1983) 628} showed that the real
part of the WRHD amplitude has a $\ln M^2_\pi$ contribution which is
computable in terms of the imaginary part of the amplitude using
dispersion relations.  For some recent work on WRHD and a more extensive
discussion of earlier work and additional references, see
ref.~\ref\singer{P. Singer, \physrev{D42}{1990}{3255}, and in Puzzles on the
Electroweak Scale, (World Scientific, 1992)}.

In this paper, we give a model independent analysis of
the WRHD parity-conserving $a$ and parity-violating $b$
decay amplitudes using chiral
perturbation theory.
Chiral perturbation theory provides the means for calculating these
decay amplitudes in terms of a systematic expansion in powers of the
Goldstone boson masses and the momentum of the radiated photon.  The
dominant contributions to the decay amplitudes can be computed in
terms of known constants with no free parameters.

The calculation of the decay amplitudes is broken down as follows.
The imaginary parts of the $a$ and $b$ decay amplitudes
are computed using unitarity as detailed by earlier authors \farrar\kogan.
The real part of the decay amplitude $b$ is dominated by a one-loop
graph which is infrared divergent in the chiral limit.  This graph
yields the $\ln M^2_\pi$ contribution discussed by Kogan
and Shifman.  The real part of $a$ is determined by pole diagrams
in addition to a one-loop
graph.  However, the computation of this graph in chiral perturbation
theory suffers from large uncertainties and the diagram cannot be
computed reliably.  In our analysis, we treat the real part of $a$
as an unknown.

All possible counterterms which contribute to WRHD are determined.
$CPS$ symmetry is used to reduce the number of counterterms.  Four
counterterms are allowed for the $a$ amplitudes, whereas only
one counterterm is allowed for the $b$ amplitudes.  The magnitude
of the $b$ counterterm
is estimated using naive dimensional analysis \ref\amhg{A. Manohar and H.
Georgi, \np{234}{1984}{189}}.
The counterterm contribution is approximately $20\%$ of
typical $b$ amplitudes.  The $b$ counterterm
does not contribute to $\spg$ or $\xmsg$ decay since the
parity violating amplitudes for these decays are purely $CPS$ violating,
as shown originally in ref.~\ref\hara{Y.~Hara, \prl{12}{1964}{378}}.

The theoretical predictions for the decay widths and
asymmetries are compared with experiment.  The
experimental data is consistent with theory with the
exception of the asymmetry parameter for $\spg$.  This
asymmetry parameter
does not agree with the data, even when $\Re a$ is treated as a
free parameter.  The experimental value of the asymmetry parameter
for $\spg$ is $\alpha =-0.83\pm0.12$.\foot{There has been a new measurement
recently by the E761 group \ref\foucher{E761 Collaboration (M.~Foucher
\etal), \prl{68}{1992}{3004}}\ of
$-0.72\pm0.086\pm0.045$. The value used in the text is that of the
1992 Particle Data Book.}
The maximum theoretical asymmetry consistent with unitarity is
$\alpha=-0.8$. It is only possible to get this upper bound value
for the asymmetry if one includes a $CPS$-violating counterterm which is
about 35 times larger than its naive value, or if the short distance
contribution is enhanced by about 20.

This paper is organized as follows.  Sect. 2 begins with a
brief description of chiral perturbation theory for baryons.
Definitions for WRHD  amplitudes, widths and asymmetries
are given.  The next three sections discuss short distance,
counterterm, and long distance contributions to WRHD amplitudes.
The short distance contribution is unimportant.  $CPS$ symmetry
is used to constrain the number of counterterms in Sect. 4.
The main computation of this paper, the calculation of long
distance contributions to the decay
amplitudes, is given in Sect. 5.  The theoretical analysis is
compared with experiment in Sect. 6.  More detailed formul\ae\
for the computation described in Sect. 5 are contained in the
appendix.

\newsec{Formalism}

The WRHD amplitudes will be computed using the static baryon formulation of
chiral perturbation theory developed in ref.~\ref\heavy{E. Jenkins
and A. Manohar, \pl{255}{1991}{558}}.  In this formalism,
baryons are described by velocity dependent fields $B_v(x)$,
where $v^\mu$ is the four-velocity of the baryon.\foot{Velocity dependent
fields
were originally introduced in the study of heavy quark symmetries in QCD
\ref\georgi{H. Georgi, \pl{240}{1990}{447}}.} The field $B_v(x)$ is related to
the
conventional baryon field $B(x)$ by the transformation
\eqn\bvfield{
B_v(x) = {1+\slash v\over 2} e^{im_B v\cdot x} B(x),
}
where $m_B$ is the baryon mass. The advantage of using the field $B_v$ is that
derivatives on the baryon field produce factors of the residual moment $k$,
which is related to the total momentum $p$ by $p=m_B v + k$.
For baryons interacting with low-momentum Goldstone bosons, the residual
momentum is small because the baryons are nearly on-shell. Higher derivative
terms in the chiral Lagrangian are then suppressed by factors of $k/\lcsb$,
where $\lcsb\sim 1$~GeV is the chiral symmetry breaking scale.
Consequently, the static baryon formulation of chiral perturbation theory
has a systematic derivative expansion.  The theory also has a systematic
loop expansion \ref\jm{E. Jenkins and
A.V. Manohar, Proceedings of the Workshop on ``Effective Field Theories
of the Standard Model'', ed. U. Meissner, World Scientific (1992)}.

The leading terms in the baryon chiral Lagrangian are
\eqn\lag{
L^0_v = i \Tr \bar B_v (v \cdot \CD) \ B_v
+ 2 D \Tr \bar B_v S_v^{\mu} \left\{ A_{\mu}, B_v \right\}
+ 2 F \Tr \bar B_v S_v^{\mu} \left[ A_{\mu}, B_v \right]
+ \ldots,
}
where
\eqn\baryon{
B_v = \pmatrix{ {1\over\sqrt2}\Sigma^0_v +
{1\over\sqrt6}\Lambda^0_v&
\Sigma^+_v & p_v\cr
\Sigma^-_v& -{1\over\sqrt2}\Sigma^0_v + {1\over\sqrt6}\Lambda^0_v&n_v\cr
\Xi^-_v &\Xi^0_v &- {2\over\sqrt6}\Lambda^0_v\cr},}
is the matrix of baryon fields,
\eqn\av{\eqalign{
&V^{\mu} = \frac 1 2 ( \xi \partial^{\mu} \xi^{\dagger}
+ \xi^\dagger \partial^{\mu} \xi ) +\frac 1 2 i e \CA^{\mu}
\left( \xi^\dagger Q \xi + \xi Q \xi^\dagger \right), \cr
&A^{\mu} = \frac i 2 (\xi \partial^{\mu} \xi^{\dagger}
  -\xi^\dagger \partial^{\mu} \xi ) - \frac 1 2 e \CA^{\mu}
\left( \xi Q \xi^\dagger - \xi^\dagger Q \xi \right),\cr
}}
$\CA^\mu$ is the photon field, and the ellipses denote terms with additional
derivatives or insertions of the light quark mass matrix. The Goldstone
bosons are contained in the field $\xi$,
\eqn\sigmaxi{
\xi = e^{i\pi/f}, \quad \Sigma = \xi^2 = e^{2 i \pi/f},
}
where
\eqn\pion{
\pi = \pmatrix{ {1\over\sqrt2}\pi^0 +
{1\over\sqrt6}\eta&
\pi^+ & K^+\cr
\pi^-& -{1\over\sqrt2}\pi^0 + {1\over\sqrt6}\eta&K^0\cr
K^- &\bar K^0 &- {2\over\sqrt6}\eta\cr
},
}
and $f \approx 132$~MeV is the pion decay constant.
Under a chiral $SU(3)_L\times SU(3)_R$ transformation,
\eqn\tran{\eqalign{
\Sigma &\rightarrow L \Sigma R^\dagger, \cr
\xi &\rightarrow L \xi U^\dagger = U \xi R^\dagger, \cr
B_v &\rightarrow U B_v U^\dagger, \cr
}}
where $U\in SU(3)$ is defined implicitly by the transformation of $\xi$.
Note that the chiral Lagrangian Eq.~\lag\ has no baryon mass term. This
feature is what ensures that the theory has a consistent loop expansion,
because large loop corrections of the form $m_B/\lcsb\sim 1$ are never
obtained since
positive powers of $m_B$
never occur in any Feynman diagram.

The effective Lagrangian for the WRHD $B_i\rightarrow B_f\gamma$ has the form
\eqn\radlag{
\CL = {e G_F\over 2}\bar B_{v\,
f}\left(a_{if}+b_{if}\gamma_5\right)\sigma^{\mu\nu}
F_{\mu\nu}B_{v \, i},
}
where $G_F$ is the Fermi coupling constant,
and $a_{if}$ and $b_{if}$ are the parity-conserving
and parity-violating decay amplitudes which have dimensions of mass.
The normalization convention used here is
that of Gilman and Wise \gw. The decay amplitude obtained from
Lagrangian~\radlag\ leads to
the decay width
\eqn\rate{
\Gamma\left(B_i\rightarrow B_f\gamma\right) = {{G_F}^2 e^2\over \pi}
\left(\abs{a}^2 + \abs{b}^2\right) \omega^3,
}
where $\omega$ is the energy of the radiated photon and
$1/m_B$ effects have been neglected. The decay angular
distribution is proportional to
\eqn\asymmetry{
{d\Gamma\over d\cos\theta} \propto 1 + \alpha\cos\theta,\quad
\alpha = {2 \Re{a b^*}\over \abs{a}^2 + \abs{b}^2},
}
where $\alpha$ is the asymmetry parameter,
and $\theta$ is the angle between the spin of the initial baryon $B_i$, and
the three-momentum of the final baryon $B_f$. For future reference,
it is convenient to define the real and imaginary parts of the decay
amplitudes, $a = a_R + i a_I$ and $b = b_R + i b_I$.

WRHD arise due to the combination of an electromagnetic interaction and a
$\ds$ weak transition. The $\ds$ Lagrangian
\eqn\dsone{
\CL^{\ds} = {G_F\over\sqrt 2}\ V_{ud}V_{us}^*\ \bar d
\gamma^\mu\left(1-\gamma_5\right) u\,
\bar u \gamma_\mu\left(1-\gamma_5\right) s,
}
is invariant under $CPS$ symmetry if mixing
to the third family is ignored.\foot{$CPS$ is $CP$ followed
by the $SU(3)$ transformation
$u\rightarrow -u$, $d\rightarrow s$, $s\rightarrow d$ which exchanges
$d$ and $s$ quarks \ref\bern{C. Bernard, T. Draper, A. Soni, H.D.
Politzer and M.B. Wise, Phys. Rev. D32 (1985) 2343}.}\ Since
$d$ and $s$ quarks have the same charge, the
electromagnetic interactions of the quarks are also invariant under $CPS$.
$CPS$ is violated in the Lagrangian because the $d$ and $s$ quarks have
different masses. Thus $CPS$ violating terms in the effective Lagrangian
will be suppressed relative to $CPS$ conserving terms by a factor of
$(m_s-m_d)/\lcsb\sim (M^2_K-M^2_\pi)/\lcsb^2\sim0.2$.

Contributions to the WRHD amplitudes can be divided into three categories:
(a) short distance contributions generated at scales large compared with
$\lcsb$,
(b) matching terms generated at the scale $\lcsb$ which are included in the
chiral
Lagrangian as local counterterms, and (c) long distance contributions that
arise from loop diagrams in the chiral Lagrangian. The precise division
of the amplitude into these
three categories is in principle scheme dependent, but it is a useful
way to organize the calculation.

\newsec{Short Distance Contributions}

The short distance contributions to WRHD are generated at energies much higher
than $\lcsb$, and can be computed using perturbation theory. The leading
operator which can produce a $s\rightarrow d\gamma$ transition
is the dimension six photonic penguin,
\eqn\penguin{
\CO = e\, G_F \ V_{ud}V_{us}^*\ \bar d \, \gamma_\mu\left(1-\gamma_5\right) s \
\partial_\nu F^{\mu\nu}.
}
However, because $\partial_\nu F^{\mu\nu}=0$ for a physical photon, this
operator
cannot contribute to WRHD. The dominant operator which can contribute to WRHD
is the quark transition magnetic moment operator,
\eqn\magnetic{
\CO= 0.2 \ { e G_F m_s\over 16\pi^2}
\ \ \bar d\,\sigma_{\mu\nu}\left(1+\gamma_5\right) s\,F^{\mu\nu}.
}
This operator is first generated by QCD radiative corrections at two loops, and
hence has a suppression factor of $\alpha_s/4\pi$. It also has a suppression
factor of a light quark mass because it is a chirality violating operator.
The numerical factor of 0.2 includes the weak mixing angles and QCD radiative
corrections \ref\svz{M.A. Shifman, A.I. Vainshtein, and V.I. Zakharov,
\physrev{D18}{1978}{2583}}. This
operator gives a contribution to the $a$ and $b$ amplitudes of order
$$
a,b\sim 0.2 \ {m_s\over 16\pi^2}
\sim  0.2 \ { M^2_K\over 16\pi^2\lcsb} \sim 0.3\ \MeV.
$$
We will find in Sect.~5 that the long distance contribution to $a$ and $b$ is
of order 5~MeV, so the short distance contribution to the decay amplitude is
negligible.
A similar conclusion has been reached previously by other authors
\kogan\singer. Note that the operator \magnetic\ is not $CPS$ violating, and
can contribute to $\spg$.

\newsec{Counterterms from Matching}

The non-perturbative
matching condition contributions to the WRHD amplitudes can be written as local
operators in the chiral Lagrangian. The magnitude and form of the counterterms
can be obtained using $SU(3)$ symmetry, $CPS$ symmetry, and naive dimensional
analysis. The WRHD counterterms contain one insertion each of the weak and
electromagnetic interactions. The $\Delta I=1/2$ enhancement implies that the
leading weak interaction operator transforms as an $SU(3)_L$ octet with the
quantum numbers of $\bar d s$. It is therefore convenient to introduce the
matrix
\eqn\hmatrix{
h = \pmatrix{ 0 &0 &0 \cr
0 &0 &1 \cr 0 &0 &0 \cr
},}
which transforms under vector $SU(3)$ as $h\rightarrow U h U^\dagger$
to describe the quantum numbers of the weak interaction. The action of the
electromagnetic interaction can be represented by the insertion of the charge
matrix
\eqn\charge{
Q = \pmatrix{ \frac 2 3 &0 &0 \cr
0 &-\frac 1 3 &0 \cr
0 &0 &-\frac 1 3 \cr},
}
which transforms under vector $SU(3)$ as $Q\rightarrow U Q U^\dagger$. We
first ignore the Lorentz structure of the counterterms, and discuss only their
flavor $SU(3)$ structure. Each possible $SU(3)$ invariant flavor structure can
contribute to either the $a$ or $b$ decay amplitudes.
The most general possible counterterms allowed by $SU(3)$ symmetry are obtained
by
considering $SU(3)$ invariant combinations of $\bar B$, $B$, $h$ and $Q$.
The tensor product of $h$ and $Q$ is ${\bf 8 \otimes 8 \rightarrow 1 +
8 + 8 + 10 + \bar {10} + 27}$. Because
$h$ and $Q$ contain only particular elements of
the complete $SU(3)$ octet, not all representations in the tensor
product are allowed. The singlet is not allowed since
$\Tr h\,Q=0$.  In addition, one octet is eliminated because $h$ and $Q$
commute, $[h,Q]=0$. Thus
$h\otimes Q \rightarrow {\bf 8 + 10 + \bar {10} + 27}$.
The initial and final baryon octets can be
combined to form ${\bf 1 + 8 + 8 + 10 + \bar {10} + 27}$. Combining these
two tensor products yields five allowed $SU(3)$ invariant
counterterms,
\eqn\cterms{\eqalign{
c_1 = \Tr\bar B\, h\, Q\, B,\quad c_2 = \Tr \bar B\, Q \, B \,h&,
\quad c_3 = \Tr \bar B\, B\, h\, Q, \quad c_4 = \Tr \bar B\, h\, Q\, B,\cr
c_5 = \Tr \bar B\, Q\ \Tr B\, h &- \Tr \bar B\, h\ \Tr B\, Q.
}}
Note that the invariant $\left( \Tr \bar B\, Q\ \Tr B\, h
+ \Tr \bar B\, h\ \Tr B\, Q \right)$ is a
linear
combination of $c_1$--$c_4$, and other possible traces can be reduced to the
above using $[h,Q]=0$.
The invariants $c_1$--$c_5$ contribute to
the WRHD amplitudes $a$ or
$b$ depending on the two possible Lorentz invariant structures of the
baryon spinor indices and the photon field
\eqn\spinorinv{
s_1 = \bar B_f \,\sigma_{\mu\nu} F^{\mu\nu} B_i,\quad
s_2 = \bar B_f \,\sigma_{\mu\nu} \gamma_5 F^{\mu\nu} B_i.
}
Thus, there are ten possible counterterms obtained by combining the five
possible flavor invariants with the two possible Lorentz invariants.

$CPS$ symmetry can be used to reduce the number of possible counterterms. All
allowed counterterms must be $CPS$ invariant when light quark masses are
neglected
since the weak and electromagnetic interactions are $CPS$ invariant. It is
straightforward to check that $CPS$ invariant counterterms must have the
flavor invariants $c_1$--$c_4$ combined with the Lorentz invariant $s_1$, or
the flavor invariant
$c_5$ combined with the Lorentz invariant $s_2$. A simple way to check $CPS$
invariance is to assume that $h$ and $Q$ are both the unit matrix. In this
case, $CPS$ symmetry reduces to $CP$.  The Lorentz invariant
$s_1$ is a $CP$ preserving magnetic dipole moment, and the Lorentz invariant
$s_2$ is a $CP$ violating electric dipole moment. Since $c_1$--$c_4$ do not
vanish when $h=Q=1$, they can only be combined with $s_1$. It takes a little
more work to show that $c_5$ (which vanishes when $h=Q=1$) is odd under $CPS$
so that it can be combined with $s_2$ to form a $CPS$ invariant counterterm.

In summary, the WRHD amplitude $a$ has four $CPS$ invariant counterterms with
flavor structure $c_1$--$c_4$, and the amplitude $b$ has only one $CPS$
invariant counterterm with flavor structure $c_5$. Note that the $c_5$
counterterm does not contribute to $\spg$ or to $\xmsg$, which was originally
shown by Hara \hara.

\newsec{Long Distance Contributions}

The long distance contribution to WRHD is obtained by computing the time
ordered product of the $\ds$ weak Lagrangian and the electromagnetic
interaction in the chiral Lagrangian.  The long distance contribution
is the dominant contribution to WRHD.

The leading $\ds$ Lagrangian is
\eqn\effdsone{
L^{\Delta S = 1}_v =
G_F M^2_{\pi^+} f \left(h_D \Tr \bar B_v \left\{ \xi^\dagger h \xi , B_v
\right\}
+  h_F \Tr \bar B_v \left[ \xi^\dagger h \xi , B_v \right] \right),
}
where the ${\bf 27}$ component of Eq.~\dsone\ has been neglected because of
the $\Delta I=1/2$ rule, and
$h_D$ and $h_F$ are dimensionless coupling constants.
The overall factor of $G_F M^2_{\pi^+}$ is included by convention.
A fit to the $s$-wave hyperon nonleptonic decays yields $h_D=0.58\pm 0.21$ and
$h_F=-1.40\pm 0.12$ \ref\nonleptonic{E. Jenkins, Nucl. Phys. B375
(1992) 561}.\foot{We use the parameter values from
tree level fit of ref.~\nonleptonic. The signs of $h_D$ and $h_F$ have been
reversed because of a different sign convention for the Lagrangian.}
Expanding Eq.~\effdsone\ in a power series in the Goldstone boson fields gives
parity conserving $\ds$ weak amplitudes of the form $G_F M^2_{\pi^+} f
\,\bar B_f
B_i$, and parity violating $\ds$ weak amplitudes of the form $G_F
M^2_{\pi^+}\,\pi \bar B_f B_i$,
\eqn\effdstwo{\eqalign{
L^{\Delta S = 1}_v =
G_F M^2_{\pi^+} f &\left(h_D \Tr \bar B_v \left\{h , B_v \right\}
+  h_F \Tr \bar B_v \left[ h , B_v \right] \right) \cr
+ i\, G_F M^2_{\pi^+} &\left(h_D \Tr \bar B_v \left\{ [h,\pi] , B_v \right\}
+  h_F \Tr \bar B_v \left[\, [h,\pi] , B_v \right] \right).
}}

The leading electromagnetic interactions are the usual interactions
proportional
to the electric charge contained in Eq.~\lag, and
the magnetic moment interaction
\eqn\lagmag{
\CL = {e \over {4 m_B}} \mu_{fi}\ \bar B_{v\,f}
\, \sigma_{\mu \nu} F^{\mu \nu} B_{v\,i}, }
where $\mu_{fi}$ is the magnetic moment in nuclear magnetons.
In the static baryon formalism, the entire baryon magnetic moment
is contained in Eq.~\lagmag, rather than just the anomalous magnetic moment.

The dominant long distance contribution to the parity-violating WRHD amplitude
$b$ comes from the one-loop diagrams shown in \fig\bampfig{One-loop
diagrams which give the leading $\ln M^2$ contribution to the
parity-violating $b$ decay
amplitudes, and to the imaginary part of the parity-conserving
$a$ decay amplitudes. The solid dots
represent strong interaction baryon-pion vertices proportional to
$D$ and $F$. Photon couplings are derived from Lagrangian $L_v^0$.
The solid squares denote the weak non-leptonic hyperon decay amplitudes. The
$s$-wave nonleptonic amplitude contributes to the $b$ WRHD amplitude, and
the $p$-wave nonleptonic amplitude contributes to the $a$ WRHD amplitude.
Graphs (e) and (f) are absent for the $b$ amplitude. Graphs with the photon
coupled to the nucleon line vanish in $v\cdot\epsilon=0$ gauge.},
where the weak vertex is the $s$-wave nonleptonic decay amplitude.
It is useful to first estimate the form of the loop diagram. The
weak vertex has the form $G_F M^2_{\pi^+} A_s$, where $A_s$ is a typical
$s$-wave nonleptonic decay amplitude and is of order one. It is a linear
combination of Clebsch-Gordan coefficients times the parameters $h_D$ and
$h_F$. The strong interaction baryon-pion vertex is $g_A k\cdot S_v/f$, where
$S_v$
is the baryon spin, $k$ is the Goldstone boson moment, and $g_A$ is a
dimensionless constant of order one which is a linear combination of $D$ and
$F$.
The Feynman integral has the form
\eqn\intform{
\int \dfour k {g_A k\cdot S_v\over f}  {G_F M^2_{\pi^+} A_s} {{e k\cdot
\epsilon}
\over (k^2)^2 k\cdot v},
}
where $k$ denotes a generic momentum, and $\epsilon$ is the polarization
vector of the photon. The gauge invariant form of the
amplitude $b$ given in Eq.~\radlag\ is $e G_F \, \omega\, \epsilon\cdot S$,
where $\omega$ is the photon energy. Thus one factor of $k$ in the loop
integral must produce a factor of the momentum of the photon, so that the
integral for $b$ has the form
\eqn\intb{
b \sim \int \dfour k {g_A \over f}  {{M^2_{\pi^+} A_s}\over k^4}.
}
This integral has an infrared divergence which is cutoff by the Goldstone boson
mass, so that
\eqn\bestimate{
b \sim {g_A M^2_{\pi^+} A_s\over 16\pi^2 f}\ln M^2/\mu^2 \sim 4\ \MeV,
}
where the numerical estimate results from
setting $A_s$ and $g_A$ to one, and using $\mu=1$ GeV. Eq.~\bestimate\ is the
$\ln M^2$ contribution discussed by Kogan and Shifman \kogan. The graph also
has an imaginary part which is determined by unitarity.

The graphs in \bampfig\ were computed in dimensional regularization, using the
methods given in ref.~\heavy, and retaining only the finite pieces. The
Goldstone
boson mass was included in the loop integral, since it regulates the infrared
divergence. Because the dominant diagrams are from pion loops, we have also
retained the $SU(3)$ mass splittings of the baryons in evaluating the loop
integrals. The  baryon propagator in the diagrams was taken to be $i/(k\cdot v
+ \Delta)$, where $k$ is the momentum along the baryon line, and $\Delta =
m-m_i$ is the difference in mass between the intermediate and initial baryons.
The graphs are most easily computed using the gauge conditions $v\cdot
\epsilon=0$ and $q\cdot \epsilon=0$ for the physical photon polarization
$\epsilon$, where
$v$ and $q$ are the velocity of the baryon and the momentum of the photon,
respectively. In $v\cdot \epsilon=0$ gauge, the only graphs
which contribute are those where the photon couples to the meson line, or at
the meson-baryon vertex. These graphs are proportional to the meson charge, so
only charged $K$ and $\pi$ loops contribute.
Both $K$ and $\pi$ loops were included in the computation, but we have checked
that the $\pi$ loops dominate. The result for the decay amplitudes is given in
Appendix~A.
The terms in the appendix include all amplitudes that do not vanish in the
$SU(3)$ limit. In addition, the contribution of the experimentally measured
$s$-wave amplitude $\Sigma^+\rightarrow n \pi^+$ is also included. This
amplitude vanishes in the $SU(3)$ limit, and is experimentally known to be
about 20 times smaller than the $SU(3)$ allowed amplitudes. Including it makes
a negligible change in our results, but we have done so to get the best
possible estimate for the imaginary part of the $b$ amplitude.

The result given in the appendix is evaluated numerically in two different
ways. The
first method (I) uses the best fit values for $h_D$, $h_F$, $D$ and $F$. The
values of $D$ and $F$ used are those given in ref.~\ref\jaffe{R.L. Jaffe
and A.V. Manohar, \np{337}{1990}{509}}.
The second method (II) uses the experimentally measured
amplitudes for $s$-wave nonleptonic decay and for the baryon semileptonic
decays to determine $A_s$ and $g_A$ wherever possible, and uses the best fit
$SU(3)$ predictions for undetermined couplings.
The results of the two methods agree with
each other, because $SU(3)$ works well for both the $s$-wave nonleptonic
amplitudes and the semileptonic decays. The $\ln\mu$ dependence completely
cancels between the $K$ and $\pi$ loops if one uses the $SU(3)$ symmetric
values for the nonleptonic decay amplitudes and axial coupling constants. In
principle there is one allowed counterterm so the graphs need not be finite.
However, the $s$-wave nonleptonic decay amplitude is not the most general
possible one allowed by $SU(3)$ symmetry, but must have the special form given
by Eq.~\effdstwo\ where $h$ and $\pi$ occur only in the combination $[h,\pi]$.
The one loop diagrams of \bampfig\ cannot produce the $SU(3)$ tensor structure
$c_5$ of Eq.~\cterms\ from $[h,\pi]$, so the graphs are finite.
The numerical values for the $b$ amplitudes are given in Table 1,
where the first and second values are obtained using methods I and II,
respectively.

In the plots in Sect. 6,  we will use method II in comparing with the
experimental data, because that gives the best approximation to the imaginary
parts of the amplitudes which are fixed by unitarity. The long distance
contribution gives an independent estimate of the local counterterm (discussed
in the previous section) to be
around $1$~MeV by looking at how much the decay amplitude changes if $\mu$ is
varied by a factor of two. Since the $K$ and $\pi$ loops together have no $\mu$
dependence, the estimate is done by looking at the variation of the $\pi$ loops
alone when $\mu$ is changed by a factor of two. This estimate is comparable to
the estimate given by naive dimensional analysis \amhg.

The computation of the $a$ amplitude is much more difficult. Naively, the
leading contribution is from the pole graphs of \fig\polefig{Pole graphs
contributing to parity-conserving $a$ decay amplitudes.  The solid
squares denote weak
$\Delta S=1$ vertices.  The photon couplings are due to baryon magnetic
moments.  },
which give
\eqn\ampa{\eqalign{
&a_{\Lambda n} ={M_\pi^2 f\over 2 m_N}\left[ -\frac 1 {\sqrt{6}} (h_D + 3 h_F)
{ {(\mu_n - \mu_\Lambda)}\over {(\Lambda - n )} }
+ \frac 1 {\sqrt{2}} (h_D - h_F) { {\mu_{\Lambda \Sigma^0}}
\over {(\Sigma^0 - n )} } \right], \cr
\noalign{\smallskip}
&a_{\Sigma^+ p} ={M_\pi^2 f\over 2 m_N}\left[ (h_D - h_F ) { {(\mu_p -
\mu_{\Sigma^+})} \over
{(\Sigma^+ - p)} }\right], \cr
\noalign{\smallskip}
&a_{\Sigma^0 n} = {M_\pi^2 f\over 2 m_N}\left[\frac 1 {\sqrt{2}} (-h_D + h_F)
{ {(\mu_n - \mu_{\Sigma^0})}\over {(\Sigma^0 - n)} }
+ \frac 1 {\sqrt{6}} (h_D + 3 h_F) { {\mu_{\Sigma^0 \Lambda}}
\over {(\Lambda - n)} }\right] , \cr
\noalign{\smallskip}
&a_{\Xi^- \Sigma^-} = {M_\pi^2 f\over 2 m_N}\left[(h_D + h_F)
{ {(\mu_{\Sigma^-} - \mu_{\Xi^-})}
\over {(\Xi^- - \Sigma^-)} } \right], \cr
\noalign{\smallskip}
&a_{\Xi^0 \Lambda} ={M_\pi^2 f\over 2 m_N}\left[ { {(\mu_\Lambda -
\mu_{\Xi^0})}
\over {(\Xi^0 - \Lambda)} } \frac 1 {\sqrt{6}} (-h_D + 3 h_F)
 -\frac 1 {\sqrt{2}} (h_D + h_F)
{ {\mu_{\Sigma^0 \Lambda}} \over {(\Xi^0 - \Sigma^0)} } \right],\cr
\noalign{\smallskip}
&a_{\Xi^0 \Sigma^0} = {M_\pi^2 f\over 2 m_N}\left[- \frac 1 {\sqrt{2}} (h_D +
h_F)
{ {(\mu_{\Sigma^0} - \mu_{\Xi^0})}
\over {(\Xi^0 - \Sigma^0)} }
+ \frac 1 {\sqrt{6}} (-h_D + 3 h_F)
{ {\mu_{\Lambda \Sigma^0}} \over {(\Xi^0 - \Lambda)} }\right] .\cr
}}
Using the best fit values for $h_D$ and $h_F$, and the known values of the
magnetic moments gives the real parts of $a$ tabulated in Table 1.
Unfortunately, the values of the pole graphs are sensitive to the precise
assumptions made to evaluate them. For example, it matters whether one uses the
leading $SU(3)$ predictions for the masses in the denominator, or the physical
values of the masses, \etc\  For the $p$-wave nonleptonic hyperon decays, there
is a cancellation between the various pole graphs, which causes the one-loop
corrections to be very important \nonleptonic. This cancellation
explains why $SU(3)$
predictions are a complete disaster for the $p$-wave nonleptonic decays.
Some cancellation also occurs between the pole diagrams for the $a$ WRHD
amplitude, so higher order corrections are expected to be important for the
$a$ amplitudes. Note that the constraints of $CPS$ do not apply to the
$a$ amplitude, because pole graphs cannot be written as local counterterms.
There is an additional complication for the $a$ amplitudes because loop graphs
of \bampfig\ with the weak vertex replaced by the nonleptonic $p$-wave
amplitude contribute. These graphs are just as important
as the $s$-wave graphs, because the nonleptonic $p$-wave amplitude
is of the same order in the derivative expansion as the $s$-wave amplitudes
\nonleptonic. These graphs cannot be computed reliably,
because the $p$-wave nonleptonic decay amplitude must be known as a
function of the pion momentum $k$, which need not be on-shell.
The typical energy scale over which the
$p$-wave amplitudes vary is of order the $SU(3)$ mass splittings in the
baryons, or of order 150 MeV, which implies that
the amplitudes are varying rapidly
in the region of interest. For these reasons, we conclude that there is no
reliable way to compute the real part of $a$. Previous work on WRHD has
produced a wide variety of estimates for the real part of $a$, which is another
indication that $a_R$ cannot be reliably calculated. In this work,
$a_R$ will be treated as an unknown parameter.

The imaginary part of $a$ can be determined reliably using unitarity. It
depends only on the $p$-wave nonleptonic decay amplitudes for on-shell pions,
which are known experimentally. We thus use the diagrams of \bampfig\ to
compute $a_I$, using the experimentally measured $p$-wave amplitudes for the
weak vertex, instead of
an $SU(3)$ fit to the $p$-wave amplitudes. The imaginary parts of the loop
graph are given in Appendix~A.
Evaluating the result numerically yields the values of $a_I$ given in Table 1.

It has been suggested recently that intermediate
spin-3/2 decuplet states should be included in chiral loop calculations
\jm\nonleptonic\ref\axialpaper{E. Jenkins and A. Manohar,
\pl{259}{1991}{353}}\ref\bmasses{E. Jenkins, \np{368}{1992}{190}}. Decuplet
intermediate states are suppressed for the $b$ amplitude, because spin-3/2
$\rightarrow$ spin-1/2 + $\pi$ $s$-wave nonleptonic decay amplitudes are
forbidden by angular momentum conservation. The real parts of $a$ do receive
contributions
from intermediate spin-3/2 states, but as we have already argued $a_R$, cannot
be computed reliably, so these contributions need not be evaluated
explicitly. There are other contributions to the real part of $a$ that we also
have not included, such as one-loop diagrams that involve the $\ds$ transition
in the meson sector from the effective Lagrangian
\eqn\meslag{
\CL^{\ds} = \lambda{f^2\over 8}\tr h D_\mu\Sigma D^\mu
\Sigma^\dagger,}
which produces the $K\rightarrow2\pi$ decay amplitude.

\newsec{Comparison with Experiment}

The results of the previous sections can now be confronted with experiment.
The imaginary parts
$a_I$ and $b_I$ can be reliably computed using unitarity. The real part $b_R$
can be reliably calculated using chiral perturbation theory. It has a typical
size of around 5 MeV, with a counterterm of typical size 1 MeV. The real part
$a_R$ cannot be computed reliably, and is treated as a free parameter.
In comparing with experiment, we use the amplitudes $a_I$, $b_I$, and $b_R$
given in Table~1, and treat $a_R$ as a free parameter. In addition, we add to
$b_R$ a counterterm contribution of the form $b_c\lambda$, where $\lambda$ is
evaluated from the $c_5$ invariant in Eq.~\cterms\ to be $-\sqrt{3/2}$,
$-3/\sqrt2$,
$\sqrt{3/2}$, and $3/\sqrt2$ for the $\lng$, $\sng$, $\xsg$, and $\xlg$ decay
modes respectively, and zero for the $\spg$ and $\xmsg$ decay amplitudes.

\nref\noble{A.J. Noble \etal, \prl{69}{1992}{414}}

The theoretical prediction is a curve in the asymmetry-amplitude plane for each
decay, as $a_R$ is varied with $a_I$, $b_R$ and $b_I$ held fixed. There is a
counterterm contribution to $b_R$, so one gets a set of curves as the
counterterm is varied. We have chosen to plot curves in
\fig\plots{Comparison of theory and experiment for the weak radiative
hyperon decays $(a)$ $\lng$, $(b)$ $\spg$, $(c)$ $\xmsg$,
$(d)$ $\xlg$, and $(e)$ $\xsg$.  The asymmetry parameter $\alpha$
and the decay amplitude $\sqrt{\abs{a}^2+\abs{b}^2}$
form the coordinate axes. The shaded
ellipses represent the experimentally allowed regions with $1 \sigma$
errors for $\alpha$ and $\Gamma$.  Theoretical curves depict the
values for $\alpha$ and $\Gamma$ obtained for a $b$ counterterm of
$0$~MeV (solid line), $1$~MeV (dashed line) and $-1$~MeV (dot-dashed
line). The unshaded regions bounded by solid curves
 for $\lng$ and $\spg$ are the recent Noble
\etal\ \noble\ and E761 measurements \foucher, respectively.}\
for $b_c=0$~MeV (solid),
$b_c=1$ MeV (dashed) and $b_c=-1$ MeV (dot-dashed). The spread in the curves
indicates the uncertainty due to the unknown counterterm $b_c$.
It is important to note
that there is only a single counterterm: one can pick either of the three kinds
of curves, but one must pick the same curve for all five decay amplitudes.

The theoretical curves in \plots\ are to be compared with the experimentally
allowed region represented by the shaded ellipse in each plot. The experimental
values used are those given in the 1992 Particle Data Book. In addition, the
recent measurements of the $\lng$ lifetime \noble, and of the $\spg$ asymmetry
parameter \foucher\ are also shown.
Keeping in mind that we expect corrections to our results of
order $m_K^2/\Lambda_\chi^2\simeq 25\%$, we see that
the results shown in \plots\ are consistent with experiment, except for the
asymmetry parameter
for $\spg$, which has been problematic for all previous work as well.
To better appreciate the significance of this disagreement between theory and
experiment, consider the $\spg$ decay with $a_R$ and
$b_R$ both treated as free parameters. The imaginary parts are determined by
unitarity, and cannot be adjusted. The maximum theoretical asymmetry for $\spg$
consistent with the observed decay width is then found to be $\alpha=-0.8$ for
$a_R=7.88\ \MeV$ and $b_R=-7.88\ \MeV$. This requires a $b$ counterterm for
$\spg$ of $-6.74\ \MeV$ to fit the observed value of $\alpha$. The naive size
of the $b$ counterterm for $\spg$ is $(m_s-m_d) (1\ \MeV)/\lcsb
\sim (M^2_K-M^2_\pi) (1\ \MeV)/\lcsb^2\sim 0.2$ MeV.
The additional suppression factor $(m_s-m_d)/\lcsb$
is present because the counterterm must be $CPS$ violating, since the $CPS$
preserving counterterm does not contribute to $\spg$. Thus the
$-6.74$~MeV counterterm required is 35 times its naive value of 0.2~MeV. The
other possibility is to have an enhancement in the short distance contribution
(discussed in Section~3) by a factor of 20.
While these possibilities cannot be ruled
out, they do  not seem likely. Thus, if the large negative asymmetry measured
for $\spg$ is correct,\foot{The asymmetry parameter for $\spg$ is
difficult to measure experimentally because of possible contamination from the
decay $\Sigma^+\rightarrow p\pi^0$, $\pi^0\rightarrow\gamma\gamma$.}\
it seems to indicate a breakdown of na\"\i ve power counting for this
process.  A possible source of this breakdown is the presence of
pole graphs containing intermediate excited $\half^-$ baryons such as the
$N^*(1535)$, which contribute to $b_R$ \poleref.

We have recently received a preprint by H. Neufeld \ref\neufeld{H. Neufeld,
UWThPh-1992-43 (1992)} which also analyzes WRHD using chiral perturbation
theory.

\bigskip\bigskip
\centerline{\bf Acknowledgements}\nobreak
This work was supported in part by the Department of Energy
under grant number DOE-FG03-90ER40546.
AVM is also supported by a National
Science Foundation Presidential
Young Investigator award number PHY-8958081.  MJS
acknowledges the support of a Superconducting Supercollider
National Fellowship from the Texas National Research Laboratory
Commission under grant FCFY9219.

\vfill\break\eject

$${\bf TABLE\ 1}: {\bf {\rm WEAK\ RADIATIVE\ HYPERON\ DECAY\ AMPLITUDES}} $$
\centerline{\vbox{
\offinterlineskip
\def\tablerule{\noalign{\hrule}}
\def\pretty{height2pt&\omit&\omit&\omit&height2pt&\omit&\omit&\omit
&height2pt&\omit&\omit&\omit&height2pt&
\omit&\omit&\omit&height2pt\cr}
\def\mystrut{\vphantom{\vrule height0.2cm}}
\def\space{&\mystrut&\omit&\omit&&&&&&&&&&&&&\cr}
\halign{
\vrule#
&\hfil\quad$#$&$#\ \rightarrow\ $&$#\gamma$\quad\hfil
&\vrule#
&\hfil$\ \ {}#$&\hfil$\ #\ $\hfil&$#\quad$\hfil
&\vrule#
&\hfil$\ \ {}#$&\hfil$\ #\ $\hfil&$#\quad$\hfil
&\vrule#
&\hfil$\ \ {}#$&\hfil$\ #\ $\hfil&$#\quad$\hfil
&\vrule#\cr
\tablerule
\pretty
\space
&\multispan3\hfil Decay\hfil&&\multispan3 \hfil$a_R + i a_I$\hfil
&&\multispan3\hfil$b_R + i b_I\ {\rm (I)}$\hfil
&&\multispan3\hfil$b_R + i b_I\ {\rm (II)}$\hfil&\cr
\space
\pretty
\tablerule
\pretty
\space
&\Lambda&&n&&-2.74&-&0.68 i&&11.75&+&11.88 i&&11.11&+&11.21 i&\cr
\space
&\Sigma^+&&p&&4.03&+&6.18 i&&-0.98&&&&-1.21&-&0.53 i&\cr
\space
&\Sigma^0&&n&&0.77&&&&5.58&+&12.66 i&&5.48&+&12.44 i&\cr
\space
&\Xi^-&&\Sigma^-&&4.62&-&1.55 i&&-7.10&-&12.16 i&&-7.26&-&12.34 i&\cr
\space
&\Xi^0&&\Lambda&&1.91&&&&-2.60&&&&-2.47&&&\cr
\space
&\Xi^0&&\Sigma^0&&-9.40&&&&2.52&&&&2.52&&&\cr
\space
\pretty
\tablerule
}}}
\smallskip
\noindent
Table 1 -- Weak Radiative Hyperon Decay Amplitudes
in units of MeV.  The two sets of values for the
$b$ amplitudes were derived using methods I and II
described in the text, respectively. The values for $a_R$ contain
only the pole graph contributions from \polefig, and are not
reliable.
\bigskip

\appendix{A}{Decay Amplitudes}
\medskip
{\leftline{\sl $b$ Amplitudes:}}

The $b$ amplitudes evaluated from the loop diagrams of \bampfig. The $s$-wave
nonleptonic decay amplitude is used at the weak vertex.
$A_s(X\bar Y \pi)$ is the $s$-wave $\ds$ weak amplitude for the process
$X\pi\leftrightarrow Y$. The amplitude is normalized to that given by
Eq.~\effdstwo, with the factor of $G_FM^2_{\pi^+}$ removed. For example,
$A_s(\Lambda \bar p\pi^+)=-(h_D+3h_F)/\sqrt 6$ is the $s$-wave nonleptonic
decay amplitude for the process $\Lambda\rightarrow p \pi^-$, which is
conventionally called $\Lambda^0_-$. The sign and normalization of the $s$-wave
amplitudes are the same as
those used by the Particle Data Group, and in Table~1 of ref.~\nonleptonic. The
coupling constant $g_A(X\bar Y\pi)$ is the baryon-Goldstone boson axial
coupling constant with the conventional normalization, $g_A(p \bar n \pi^-) =
D+F \sim 1.26$, \etc\ $I_1$ and
$I_2$ are integrals that are given in Appendix~B. The arguments of the
integrals are the masses of the particles with the given labels.

\eqn\answerb{\eqalign{
b_{\Lambda n} &=
A_s(\Lambda\bar p\pi^+) g_A(p\bar n\pi^-)
I_1(\Lambda,p,n,\pi)\cr
&-A_s(\Sigma^-\bar n\pi^+)g_A(\Lambda\bar\Sigma^-\pi^-)
I_2(\Lambda,\Sigma^-,n,\pi)
\cr
&+A_s(\Sigma^+\bar n \pi^-) g_A(\Lambda\bar\Sigma^+\pi^+)
I_2(\Lambda,\Sigma^+,n,\pi)\cr
&-A_s(\Lambda\bar\Sigma^-K^-) g_A(\Sigma^-\bar n K^+)
I_1(\Lambda,\Sigma^-,n,K)
\cr
&+A_s(p\bar n K^-)g_A(\Lambda\bar p K^+)
I_2(\Lambda,\Sigma^-p,n,K)
\cr
}}
\eqn\ansbsp{\eqalign{
b_{\Sigma^+ p} &=
 -A_s(\Lambda\bar p\pi^+)g_A(\Sigma^+\bar\Lambda\pi^-)
I_2(\Sigma^+,\Lambda,p,\pi)\cr
&-A_s(\Sigma^0\bar p\pi^+)g_A(\Sigma^+\bar\Sigma^0\pi^-)
I_2(\Sigma^+,\Sigma^0,p,\pi)
\cr
&-A_s(\Sigma^+\bar n \pi^-) g_A(n\bar p \pi^+)
I_1(\Sigma^+,n,p,\pi)\cr
&-A_s(\Sigma^+\bar\Sigma^0 K^-)g_A(\Sigma^0 \bar p K^+)
I_1(\Sigma^+,\Sigma^0,p,K)
\cr
&-A_s(\Sigma^+\bar \Lambda K^-) g_A(\Lambda \bar p K^+)
I_1(\Sigma^+,\Lambda,p,K)
\cr
}}
\eqn\ansbsn{\eqalign{
b_{\Sigma^0 n} &=
A_s(\Sigma^0 \bar p \pi^+) g_A(p \bar n \pi^-)
I_1(\Sigma^0,p,n,\pi)\cr
&-A_s(\Sigma^-\bar n \pi^+) g_A(\Sigma^0 \bar \Sigma^- \pi^-)
I_2(\Sigma^0,\Sigma^-,n,\pi)\cr
&+A_s(\Sigma^+\bar n \pi^-) g_A(\Sigma^0\bar\Sigma^+\pi^+)
I_2(\Sigma^0,\Sigma^+,n,\pi)\cr
&-A_s(\Sigma^0 \bar\Sigma^- K^-) g_A(\Sigma^- \bar n K^+)
I_1(\Sigma^0,\Sigma^-,n,K)\cr
&+A_s(p \bar n K^-) g_A(\Sigma^0 \bar p K^+)
I_2(\Sigma^0,p,n,K)
\cr
}}
\eqn\ansbxs{\eqalign{
b_{\Xi^-\Sigma^-} &=
A_s(\Xi^-\bar \Lambda \pi^+)g_A(\Lambda\bar \Sigma^-\pi^-)
I_1(\Xi^-,\Lambda,\Sigma^-,\pi)\cr
&+A_s(\Xi^- \bar \Sigma^0 \pi^+) g_A(\Sigma^0 \bar \Sigma^- \pi^-)
I_1(\Xi^-,\Sigma^0,\Sigma^-,\pi)\cr
&+A_s(\Lambda \bar\Sigma^- K^-)g_A(\Xi^-\bar \Lambda K^+)
I_2(\Xi^-,\Lambda,\Sigma^-,K)\cr
&+A_s(\Sigma^0 \bar\Sigma^- K^-) g_A(\Xi^-\bar \Sigma^0 K^+)
I_2(\Xi^-,\Sigma^0,\Sigma^-,K)
\cr
}}
\eqn\ansbxl{\eqalign{
b_{\Xi^0\Lambda} &=
A_s(\Xi^0 \bar\Sigma^+\pi^+) g_A(\Sigma^+\bar\Lambda \pi^-)
I_1(\Xi^0,\Sigma^+,\Lambda,\pi)\cr
&-A_s(\Xi^-\bar\Lambda\pi^+) g_A(\Xi^0\bar\Xi^-\pi^-)
I_2(\Xi^0,\Xi^-,\Lambda,\pi)\cr
&-A_s(\Xi^0 \bar\Xi^- K^-) g_A(\Xi^- \bar\Lambda K^+)
I_1(\Xi^0,\Xi^-,\Lambda,K)\cr
&+A_s(\Sigma^+\bar\Lambda K^-) g_A(\Xi^0\bar\Sigma^+ K^+)
I_2(\Xi^0,\Sigma^+,\Lambda,K)
\cr
}}
\eqn\ansbxsz{\eqalign{
b_{\Xi^0\Sigma^0} &=
A_s(\Xi^0\bar\Sigma^+\pi^+) g_A(\Sigma^+\bar\Sigma^0\pi^-)
I_1(\Xi^0,\Sigma^+,\Sigma^0,\pi)\cr
&- A_s(\Xi^-\bar\Sigma^0\pi^+) g_A(\Xi^0\bar\Xi^-\pi^-)
I_2(\Xi^0,\Sigma^-,\Sigma^0,\pi)\cr
&-A_s(\Xi^0\bar\Xi^- K^-) g_A(\Xi^- \bar\Sigma^0 K^+)
I_1(\Xi^0,\Xi^-,\Sigma^0,K)\cr
&+A_s(\Sigma^+\bar\Sigma^0 K^-) g_A(\Xi^0\bar\Sigma^+ K^+)
I_2(\Xi^0,\Sigma^+,\Sigma^0,K)\cr
}}
\medskip
\leftline{{\sl $a$ Amplitudes:}}
The imaginary parts of the $a$ amplitudes evaluated from the loop diagrams of
\bampfig. The measured values of the $p$-wave nonleptonic decay amplitudes are
used at the weak vertices.
The $p$-wave amplitudes have the normalization used in Table~1 of
Ref.~\nonleptonic.
\eqn\ampaim{\eqalign{
&\ai_{\Lambda n} = -A_p(\Lambda \bar p \pi^+)g_A(p \bar n
\pi^-)J(\Lambda,p,n,\pi),\cr
&\ai_{\Sigma^+ p} = A_p(\Sigma^+\bar n \pi^-)g_A(n \bar p
\pi^+)J(\Sigma^+,n,p,\pi) ,\cr
&\ai_{\Sigma^0 n} = 0,\cr
&\ai_{\Xi^- \Sigma^-} = -A_p(\Xi^- \bar \Lambda \pi^+)
g_A(\Lambda \bar \Sigma^- \pi^-) J(\Xi^-,\Lambda,\Sigma^-,\pi),\cr
&\ai_{\Xi^0 \Lambda} = 0,\cr
&\ai_{\Xi^0 \Sigma^0} = 0,\cr
}}
where the integral $J$ is defined in Appendix~B.

\appendix{B}{Integrals}

\eqn\ia{\eqalign{
I_1(m_i,m,m_f,M) &=\Biggl[{M_{\pi^+}^2\over 8 \pi^2 f\omega}\Biggr]
\Biggl[\half\omega\,\left( 2 - \ln (M^2/\mu^2)\right)\cr
&\quad+2 \int_0^1 dx \, f_1(\Delta-\omega x,M) - 2 f_1(\Delta,M)\Biggr],
\cr
}}
where
$$
\Delta = m_i-m, \qquad \omega=m_i-m_f,
$$

\eqn\fa{
f_1(y,M) = \cases{\sqrt{M^2-y^2}\left[{\pi\over 2}
+ \tan^{-1}\left( {y\over
\sqrt{M^2-y^2}}\right)\right],&$\abs{y} \le M$,\cr
\half\sqrt{y^2-M^2}\left[-2 i \pi + \ln \left({y +\sqrt{y^2-M^2}\over
y-\sqrt{y^2-M^2}}\right)\right],&$\abs{y} > M$.\cr
}}

\eqn\ib{\eqalign{
I_2(m_i,m,m_f,M) &= \Biggl[{M_{\pi^+}^2\over 8 \pi^2 f\omega}\Biggr]
\Biggl[\half\omega \left( 2 - \ln (M^2/\mu^2) \right)\cr
&\quad - 2 \int_0^1 dx  \, f_2(\Delta+\omega x, M) + 2
f_2(\Delta,M)\Biggr],
}}
where
$$
\Delta = m_f-m, \qquad \omega=m_i-m_f,
$$

\eqn\fb{
f_2(y,M) = \cases{\sqrt{M^2-y^2}\left[{\pi\over2}
+ \tan^{-1}\left({y\over
\sqrt{M^2-y^2}}\right)\right],&$\abs{y} \le M$,\cr
\half\sqrt{y^2-M^2}\left[\ln \left({-y -\sqrt{y^2-M^2}\over
  -y+\sqrt{y^2-M^2}}\right)\right],&$\abs{y} > M$.\cr
}}

\eqn\jfunc{
J(x,y,z,M) = {M^2_{\pi^+}\over 8 \pi f \omega^2}
\left[\Delta \sqrt{\Delta^2-M^2} + M^2
\ln \left({M\over \Delta +
\sqrt{\Delta^2-M^2}}\right)\right],
}
where
$$
\Delta=x-y,\quad \omega=x-z.
$$

\listrefs
\listfigs
\insertfig{Figure 1}{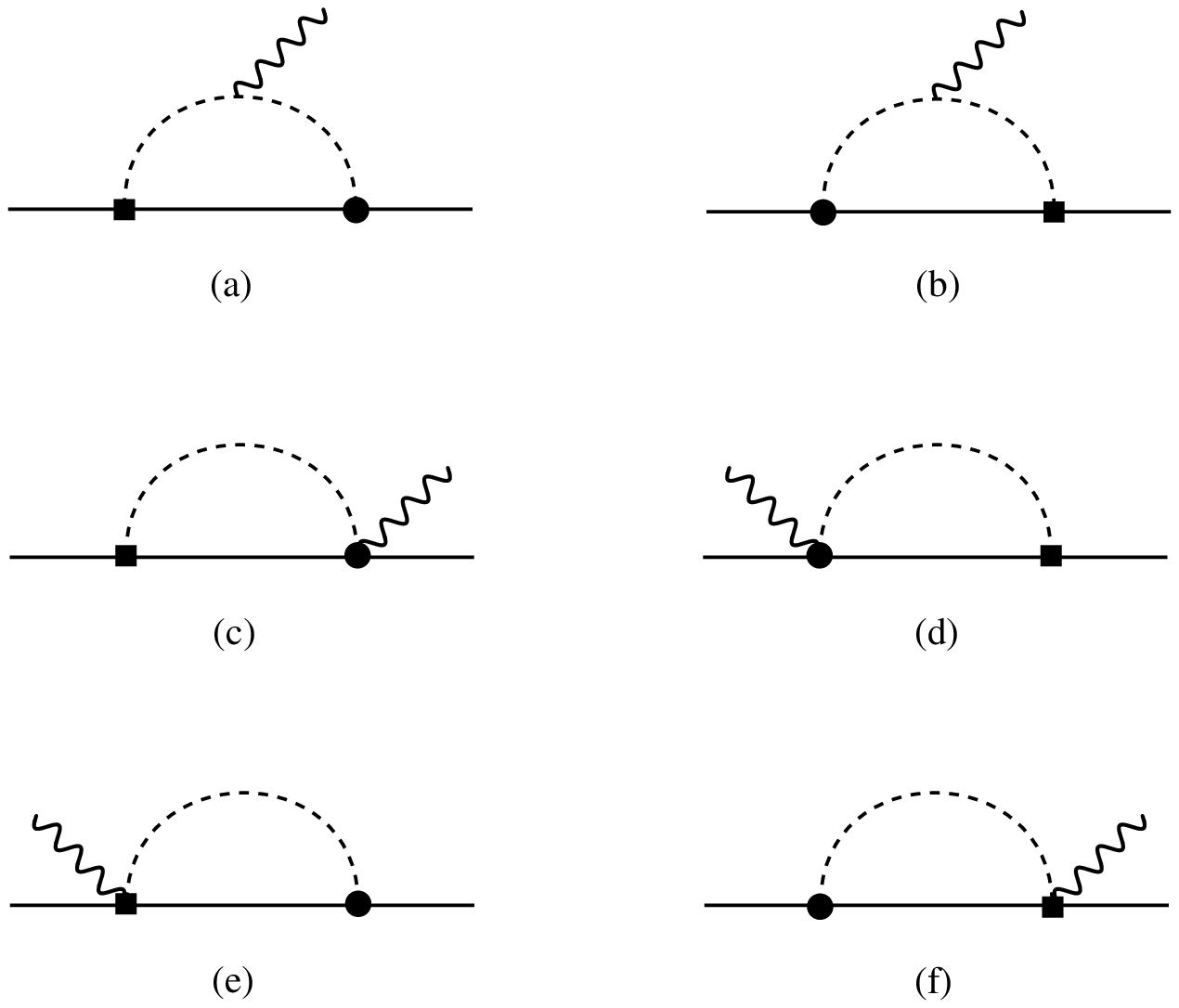}
\insertfig{Figure 2}{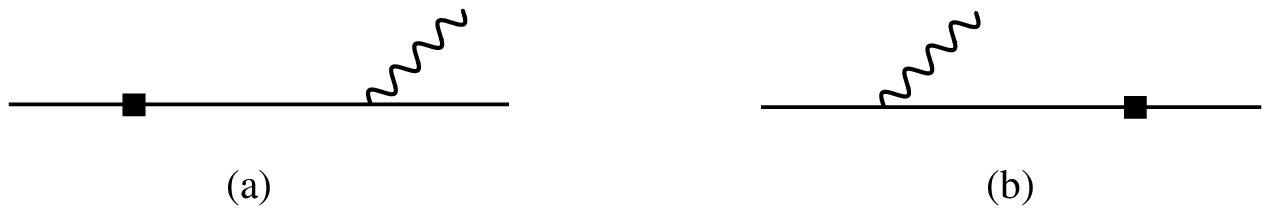}
\insertplot{(a) $\Lambda\rightarrow n\gamma$}{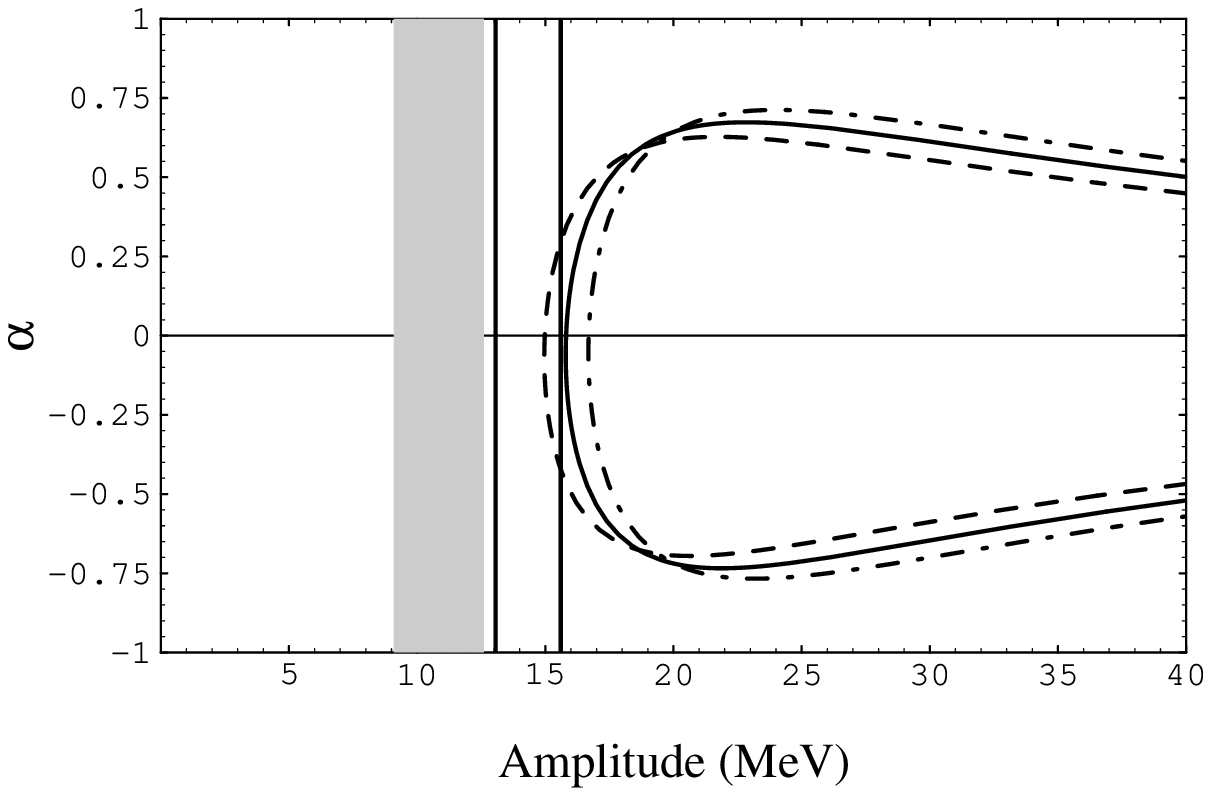}
\insertplot{(b) $\Sigma^+\rightarrow p\gamma$}{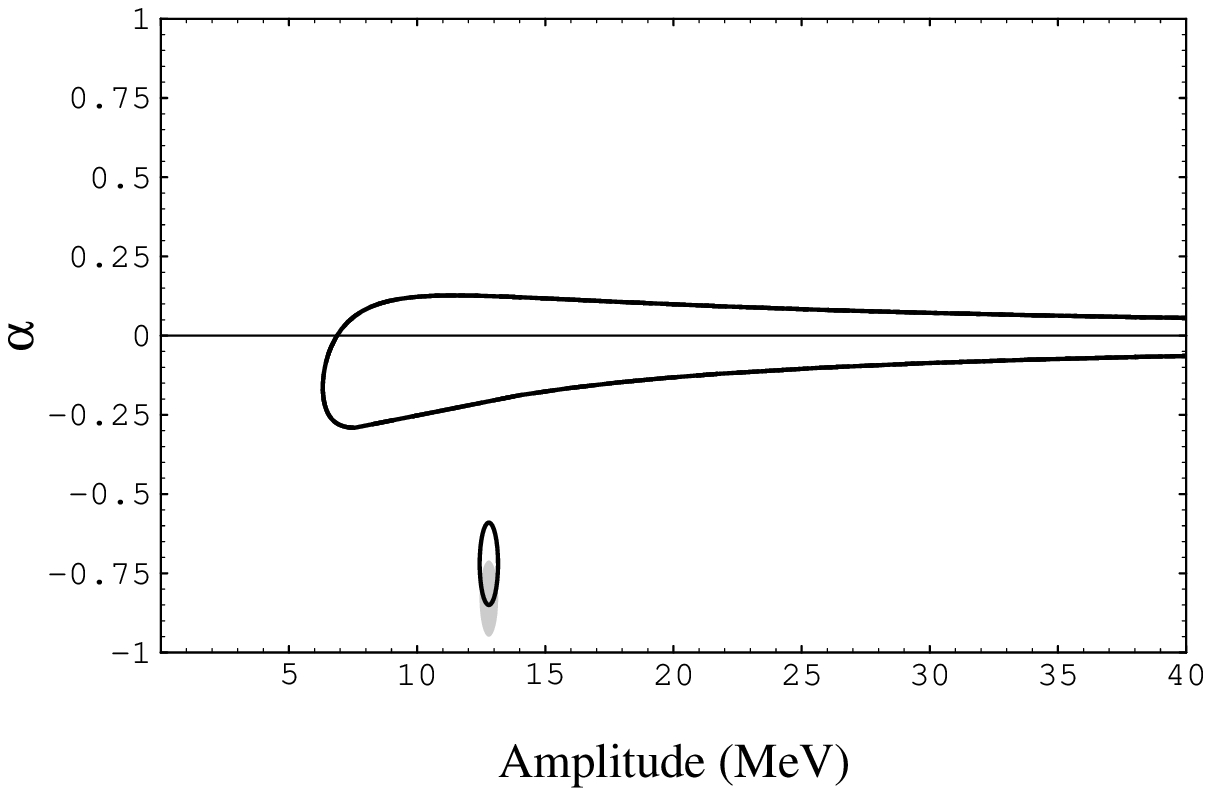}
\insertplot{(c) $\Xi^-\rightarrow \Sigma^-\gamma$}{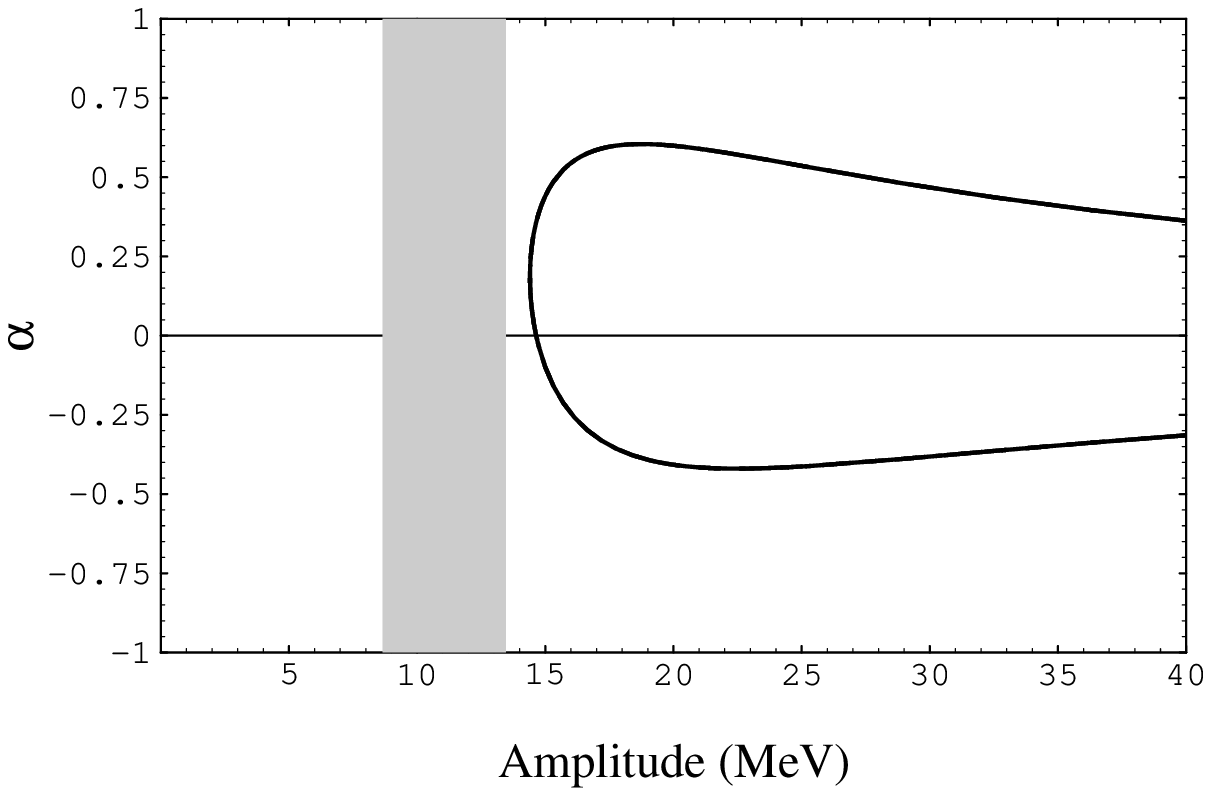}
\insertplot{(d) $\Xi^0\rightarrow \Lambda\gamma$}{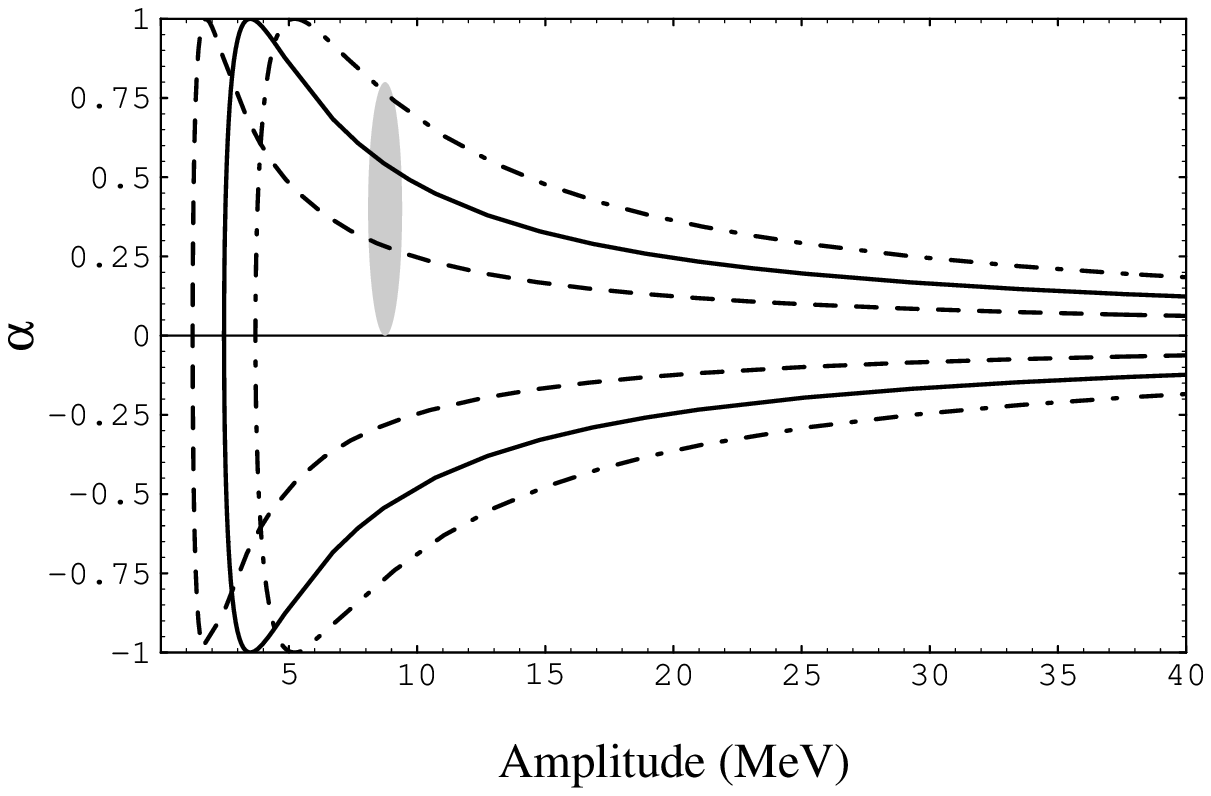}
\insertplot{(e) $\Xi^0\rightarrow \Sigma^0\gamma$}{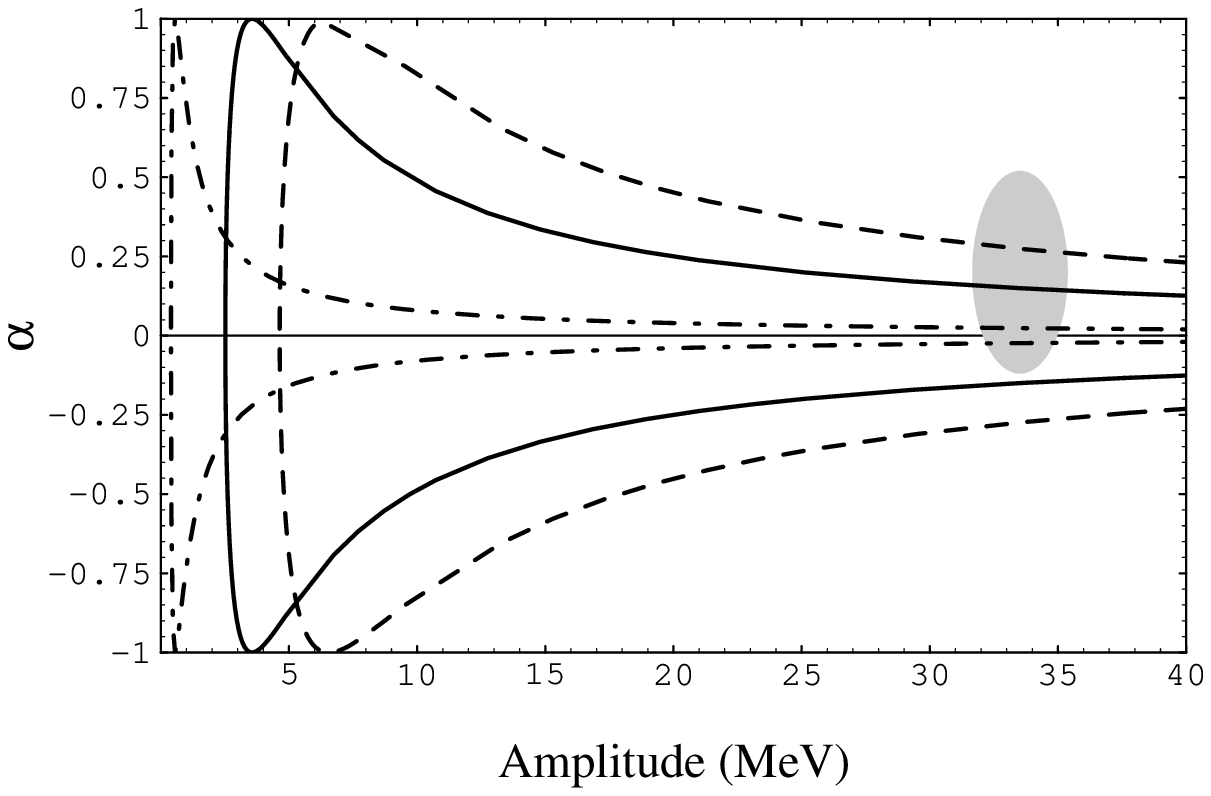}
\bye